\documentclass[fleqn,twoside]{article}
\usepackage[headings]{espcrc2}

\readRCS
$Id: espcrc2.tex,v 1.2 2004/02/24 11:22:11 spepping Exp $
\usepackage{graphicx}
\usepackage[figuresright]{rotating}

\newcommand{\AmS}{{\protect\the\textfont2
  A\kern-.1667em\lower.5ex\hbox{M}\kern-.125emS}}

\newcommand{\ppbar}{$\overline{p}p$}

\newcommand{\ep}{$e^+e^-$}

\newcommand{\psipiz}{$J/\psi\pi^0$}
\newcommand{\etacgam}{$\eta_c \gamma$}

\newcommand{\piz}{$\pi^0$}

\newcommand{\ra}{$\rightarrow$}
\title{E835 at FNAL: Charmonium Spectroscopy in $\bar p p$  Annihilations}

\author{Claudia Patrignani\address{
Istituto Nazionale di Fisica Nucleare and University of Genova, 
        Via Dodecaneso 33, \\ 
        16146 Genova, Italy}
        (FNAL-E835 Collaboration)%
\thanks{
M.Andreotti$^{2}$, S.Bagnasco$^{3,7}$, W.Baldini$^{2}$, D.Bettoni$^{2}$,
G.Borreani$^{7}$, A.Buzzo$^{3}$, R. Calabrese$^{2}$, R. Cester$^{7}$,
G.Cibinetto$^{2}$, P.Dalpiaz$^{2}$, G.Garzoglio$^{1}$, K.E.Gollwitzer$^{1}$,
M.Graham$^{5}$, M.Hu$^{1}$, D.Joffe$^{6}$, 
J.Kasper$^{6}$, G.Lasio$^{4}$, M.Lo~Vetere$^{3}$, E.Luppi$^{2}$, 
M.Macr\`\i$^{3}$, M.Mandelkern$^{4}$, F.Marchetto$^{7}$, M.Marinelli$^{3}$, 
E.Menichetti$^{7}$, Z.Metreveli$^{6}$, R.Mussa$^{7}$, M.Negrini$^{2}$, 
M.M.Obertino$^{7}$, M.Pallavicini$^{3}$, N.Pastrone$^{7}$, 
T.K.Pedlar$^{6}$, S.Pordes$^{1}$, E.Robutti$^{3}$, W.Roethel$^{6,4}$, 
J.L.Rosen$^{6}$, P.Rumerio$^{6}$, R.Rusack$^{5}$, A.Santroni$^{3}$, 
J.Schultz$^{4}$, S.H.Seo$^{5}$, K.K.Seth$^{6}$, G.Stancari$^{1}$, 
M.Stancari$^{4}$, A.Tomaradze$^{6}$, I.Uman$^{6}$, T.Vidnovic~III$^{5}$, 
S.Werkema$^{1}$ and P.Zweber$^{6}$ ~~~~~~~~~~~~~~~~~~~~~~~~~~~~~~~~~~~~~~~~~~~~~~~~~~~~~~~
\hbox{$^{1}$Fermilab, Batavia, Illinois 60510,} ~~~~~~~~~~~~~~~~~~~~~~~~~~~~~~~~~~~~~~~~~~~~~~
\hbox{$^{2}$I.N.F.N. and University of Ferrara,} ~~~~~~~~~~~~~~~~~~~~~~~~~~~~~~~~~~~~~~~~~~~~~~
\hbox{$^{3}$I.N.F.N. and University of Genova,} ~~~~~~~~~~~~~~~~~~~~~~~~~~~~~~~~~~~~~~~~~~~~~~~
\hbox{$^{4}$University of California at Irvine,} ~~~~~~~~~~~~~~~~~~~~~~~~~~~~~~~~~~~~~~~~~~~~~~~~
\hbox{$^{5}$University of Minnesota,~~~~~~~~} ~~~~~~~~~~~~~~~~~~~~~~~~~~~~~~~~~~~~~~~~~~~~~~~~~~~~~~~~~~~~~~~~
\hbox{$^{6}$Northwestern University, Evanston,IL,~~~~~~~} ~~~~~~~~~~~~~~~~~~~~~~~~~~~~~~~~~~~~~~~~~~~~~~~~~~~~~~~~~~~~~~
\hbox{$^{7}$I.N.F.N. and University of Torino.}}}

\runauthor{C. Patrignani}

\begin{document}

\begin{abstract}
I present preliminary results on the search for $h_c$ in its \etacgam\ and
\psipiz\ decay modes.
We observe an excess of \etacgam\ events
near 3526 MeV that has a probability
${\cal P} \sim 0.001$ to arise from background fluctations. 
The resonance parameters are $M=3525.8 \pm
0.2 \pm 0.2\,$MeV, $\Gamma\leq$ 1 MeV, and 
$10.6\pm 3.7\pm3.4(br) < \Gamma_{\overline{p}p}B_{\eta_c\gamma} < 12.8\pm 4.8\pm4.5(br)\,$eV.
We find no event excess within the search region in the \psipiz~mode.
\vspace{1pc}
\end{abstract}

\maketitle

\section{INTRODUCTION}
Charmonium states have been succesfully studied in \ppbar\ annihilations,
where all states can be formed, and detected in their decay to electromagnetic
final states. E760 and E835 measured precisely masses and 
widths of $\eta_c$, $J/\psi$, $\psi(2S)$ and $\chi_{cJ}$ states
as well as other properties for these states, e.g., $\Gamma_{\gamma\gamma}$ and 
${\cal B}_{p\bar p}$\cite{chi0}.

The $h_c$ is so far the most elusive
charmonium state below ${\bar D} D$ threshold~\cite{Barnes}.
It has not been observed yet in $\psi(2S)$ decays, where it could be
produced by sequential radiative
transitions through the $\chi_{c2}$ (E1 followed by M1), or by 
I~spin violating $\pi^0$ transition, nor in B decays whose branching ratios 
to $h_c$ could be as large as   
${\mathcal{O}}(10^{-4})\,$\cite{Bdec}.

Observation of the $h_c(1^1P_1(1^{+-}))$  
will allow to determine the splitting between singlet P  and 
the spin-weighted average mass of the triplet P
states ($\chi_{c.o.g}=3525.30\pm0.07\,$MeV~\cite{pdg}).
In one-gluon potential model, this splitting is zero.   
Corrections are small and the mass difference between the $h_c$ 
and the $\chi_{c.o.g}$
is in general predicted to be at most a few MeV~\cite{theory}, thus
it is important to measure the $h_c$ mass to better than $\approx1\,$MeV.

The $h_c$ is expected to be narrow ($<1\,$MeV in width),
to have a dominant $E1$ transition to \etacgam,
and large branching ratio to light hadrons~\cite{bodwin:1994jh}.

Our study of $\chi_{c0}$ has shown that 
E1 transitions of triplet P states are in 
excellent agreement\cite{pdg} with the predicted scaling as 
the third power of photon momentum. 
Assuming that the $h_c$ has 
the same radial wave function as the other P states, this
would imply
$\Gamma(h_c\to\eta_c\gamma)\approx 600\,$keV 
for an $h_c$ mass close to the $\chi_{c.o.g}$. 

In 1992, E760 (our former experiment) reported observation of a structure 
in the cross section \ppbar\ra\psipiz\ (an I-violating mode) close 
to the $\chi_{c.o.g.}$ interpreted as the $h_c\,$\cite{armstrong:1992}. 
Neglecting interference with continuum
\ppbar\ra\psipiz, resonance parameters
were determined as $M=3526.2 \pm 0.15 \pm
0.2\,$MeV; $\Gamma \leq 1.1\,$MeV (90\% CL);
and $(1.8\pm0.4)\cdot10^{-7}< {\cal B}(p\bar p){\cal B}(J/\psi\pi^0)<(2.5\pm0.6)\cdot 10^{-7}$.
The probability for such structure to arise
from background was estimated to be 1/400. 

Observation of the $h_c$
is one of the principal objectives of experiment
E835 at Fermilab.

\section{EXPERIMENT E835}
Charmonium states are studied by 
a scan of the
$\bar p p$ annihilation cross section for exclusive  final 
states at different center of mass 
energies ($E_{CM}$).  
An excess of events at any value of 
${E}_{CM}^*$ over the background measured on a broader $E_{CM}$
range signals the formation of a resonance.
Resonance parameters are then determined with 
precision up to $~100\,$keV on masses and widths.
The experiment is designed
to observe charmonium states in their decays to electromagnetic final states such as
$J/\psi\,X\to e^+e^-\,X$ and $\eta_c(\gamma)\to \gamma\gamma(\gamma)$.

E835 is a major upgrade of  E760 and is decribed in detail 
in~\cite{bigpaper}. 
The detector is a non-magnetic, large acceptance,
cylindrical spectrometer, covering the complete azimuth ($\phi$) and
from $2^{\circ}$ to $70^{\circ}$ in polar angle ($\theta$).
It consists of a lead glass electromagnetic calorimeter divided into
a barrel and a forward sections; the inner part of the barrel section
is instrumented with a multicell threshold \v{C}erenkov counter, for electron
detection, three
concentric scintillator hodoscopes,  and a tracking system to measure 
charged particles. To withstand the
$\sim 5$ MHz continuous interaction rate, all channels are
instrumented with multi-hit TDCs.

The required high luminosity is achieved by
a $H_2$ jet target intersecting the
$\bar p$ beam in the Fermilab Antiproton Accumulator.
The beam is decelerated from the accumulation energy 
to the value appropriate for the formation of each resonance.
The density of the target is continuously adjusted
to compensate for beam loss
keeping the instantaneous luminosity constant at $\sim 2\times10^{31}$
cm$^{-2}$s$^{-1}$.  
The stochastic cooling keeps beam momentum
constant with a $\Delta$p/p$ \approx 2\times10^{-4}$,
compensating for energy losses in the target.

The integrated luminosity $L_{int}$ for each energy setting is measured to 
$<2\%$ by counting
recoil protons from $\bar{p}p$ scattering at $\theta \approx 90^{\circ}$.

\section{SEARCH FOR $h_c\to\eta_c\gamma$ and $J/\psi\pi^0$}

E835 took data in 1996/1997 (E835-I)
for an integrated luminosity $L_{int}\approx 140\,$pb$^{-1}$,
and again in 2000 (E835-II) for
$L_{int}\approx 110\,$pb$^{-1}$.

We search for $h_c$ in the following reactions
\begin{equation}
{\bar p} p  \rightarrow  h_c  \rightarrow  J/\psi \pi^0\to e^+e^-\gamma\gamma
\label{psipio}
\end {equation}
\begin{equation}
{\bar p} p  \rightarrow  h_c  \rightarrow  \eta_c \gamma\to 3\gamma
\label{tregam} 
\end{equation}
The scan for the $h_c$ was based
on assumptions for the mass (close to $\chi_{c.o.g.}$), total width ($<$1~MeV)
and expected yields at resonance peak: 3-8~ev/pb$^{-1}$ (above a continuum
yield of $\approx\,$2~ev/pb$^{-1}$) for reaction (\ref{psipio})
and $<1\,$ev/pb$^{-1}$ for reaction (\ref{tregam})~\cite{explan}.

We have taken data for about $215\,$pb$^{-1}$ in the range 
$3300<E_{CM}<4400\,$MeV. Approximately $70\,$pb$^{-1}$ were spent
in a fine scan of the $\chi_{c.o.g.}$ region $3525.2<E_{CM}<3527.2\,$MeV and
$20\,$pb$^{-1}$ a coarser scan of $3520<E_{CM}<3540\,$MeV 
(see Fig.~\ref{lumplot}), while the
remaining data are used to measure background.
Data taken at $\chi_{c1}$ and $\chi_{c2}$, on which we performed
repeated scans which will provide new measurements of their masses
and widths, provide also clean $J/\psi\gamma$ events to
monitor efficiencies and nearby background measurement for
reaction~(\ref{tregam}) on either side of the $\chi_{c.o.g}$. 




\subsection{Search for $\bar p p\to h_c\to$\etacgam$\to 3\gamma$}

The analysis is based on the study of simulated $h_c$ events and a background
sample of $p\bar p\to 3\gamma$ candidates (prior to the $\eta_c$ mass 
constraint) from a fraction of data taken ouside the $\chi_{c.o.g}$, 
then counting $\eta_c\gamma$ candidates in the whole sample.

Candidates for neutral final states are selected by the 
trigger if there are no charged tracks from the interaction point
and there are at least 2 energy deposits in the 
Central Calorimeter (CCAL) with invariant mass 
$\geq 2.2\,$GeV, or if the energy detected in CCAL 
is $>$80\% of the total energy.

\begin{figure}[htb] \begin{center} 
\includegraphics[scale=0.4]{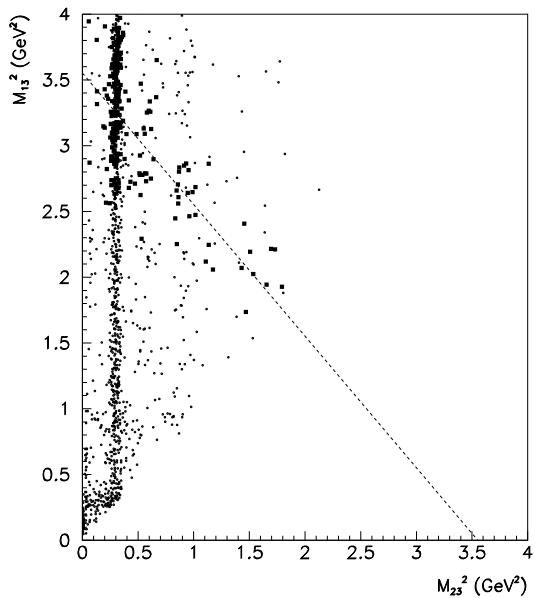}
\end{center} 
\caption{Dalitz plot for candidates \ppbar\ra$3\gamma$ events in the 
$\cos{\theta^*}$ acceptance 
described in the text
(small dots) for $3524< E_{CM} <3527$ MeV and 
\ppbar\ra\etacgam\ candidates (large squares).
The dotted line shows the center of the \etacgam\ band. \ppbar\ra\etacgam 
candidates with $M_{23}>1\,$GeV have relatively little background.
{\it ~(Preliminary)}}
\label{dalitz1} \end{figure}

\begin{figure}[htb] 
\begin{center}
\includegraphics[scale=0.4]{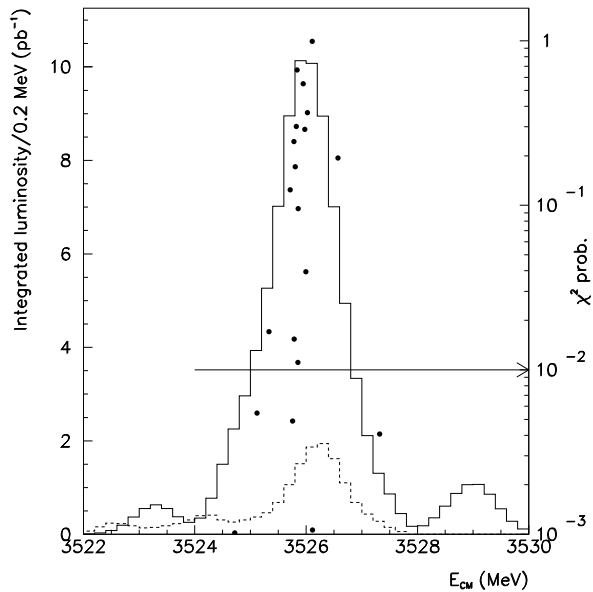}
\end{center}
\caption{E835 integrated luminosity vs $E_{CM}$ 
in the $\chi_{c.o.g.}$
region. Dashed histogram is the corresponding E760 luminosity.
Superimposed (dots, scale on right) we show the $\chi^2$ probability 
distribution vs $E_{CM}$
for $\eta_c\gamma$ candidates. Events with large probabilities
are observed in a narrow $E_{CM}$ range of our high-luminosity scan. Arrow indicates the
probability cut applied. {\it ~(Preliminary)}}
\label{lumplot} \end{figure}
Offline we select events with 3 ``on-time'' {\it candidate} photons in CCAL, 
defined as clusters with energy $>20\,$MeV occuring
within 6 ns of the trigger. We reject events with on-time clusters in the forward calorimeter.
Since timing efficiency and resolution are worse for low energy 
photons,
clusters without timing information, or with $E<300$ MeV and
occuring within 6 and 15 ns of the trigger are considered {\it undetermined}. 
If a candidate photon paired with any other candidate or
undetermined cluster forms a $\pi^0$ ($|m_{\gamma\gamma}-m_{\pi^0}|<35$ MeV)
the event is rejected.   
Events with signals in the two outer hodoscopes not in
coincidence with the corresponding Cerenkov counter are rejected; those with
coincidences are retained as events in which a $\gamma$ converted after 
the innermost hodoscope.
We impose a likelihood ratio test (PW) on photon showers analogous to the 
electron weight (EW) 
described in Ref.~\cite{bigpaper}, but
based only on CCAL cluster moments.
We require $PW_1\times PW_2>1$ and
$PW_3>1.5$, where the the photons are ordered by
their CM energies:
$E_{\gamma1}>E_{\gamma2}>E_{\gamma3}$
The efficiency of this cut on clean
$J/\psi \gamma \rightarrow e^+e^-\gamma $ events at $\chi_{c1}$
and $\chi_{c2}$, of energies comparable to those of $h_c$ 
radiative decay, is constant and it is well modeled
by the simulation.

A 4C kinematic fit to the hypothesis
\ppbar\ra$3\gamma$ is performed and we require a nominal $\chi^2$
probability {\cal P}$(3\gamma)>10^{-4}$. If there are undetermined
clusters, we require that
{\cal P}$(3\gamma)>${\cal P}$(4\gamma)$, the latter being the 
probability for any fit to \ppbar\ra$4\gamma$.


\begin{figure}[htb] \begin{center} 
\includegraphics[scale=0.4]{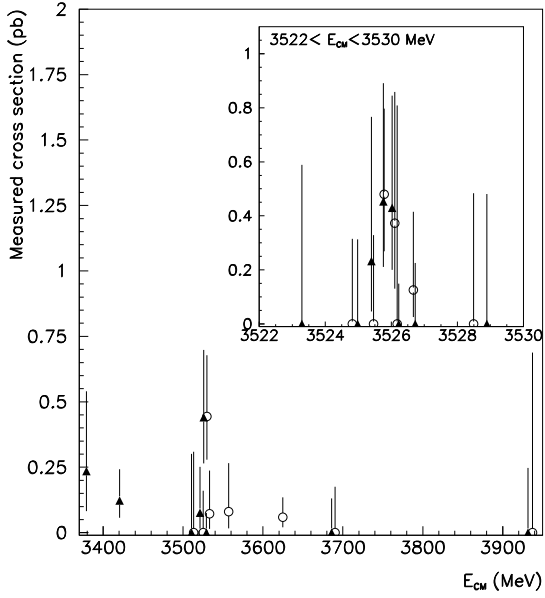}
\end{center} 
\caption{Cross section for \ppbar\ra\etacgam\ra$\gamma\gamma\gamma$.  
The inset is an expanded view of the $\chi_{c.o.g}$ region. E835-I
data are open circles and E835-II data closed triangles. {\it ~(Preliminary)}}
\label{cross} \end{figure}
$3\gamma$ candidates have background mostly from
\ppbar\ra$\pi^0\eta$, $\pi^0\eta'$, and 
$3\pi^0$ events, where the $\pi^0$s decay symmetrically or 
highly asymmetrical; for such  $\pi^0$'s
only one cluster, in approximately the direction of the
$\pi^0$, is detected and identified as a $\gamma$. We refer to such events as
as {\it feed-down} events.
$\bar p p$ annihilations to two and three light mesons are strongly 
forward/backward-peaked and 
high energy photons  
from feed-down events tend to have large $\vert\cos{\theta^*_{1,2}}\vert$. 
\ppbar\ra\piz$X^0$ reactions are forward-backward symmetric,
but forward \piz\ are more likely 
to be misidentified as single photons, thus 
feed-down events have often low energy $\gamma\,$s emitted 
backwards, with invariant mass of the two lowest energy candidates 
$M_{23}\approx M_X$.

For reaction (\ref{tregam})
the angular distribution is nearly uniform in $\cos{\theta^*_{12}}$ for the two photons
from $\eta_c$ and $\propto
\sin^2{\theta^\ast_3}$ for the radiative decay photon. 

Based on signal and background characteristic we cut at
$M_{23}>1\,$GeV, $|\cos \theta^\ast_{1,2}|<0.5$ and
$-0.4<\cos{\theta^\ast}_{3}<0.7$.

We apply a 5C kinematical fit to \etacgam\ and set a
$\chi^2$ probability cut at $>0.01$.
The overall efficiency of this selection is 3.2\%.

\begin{table}[htb]
\begin{center}
\renewcommand{\arraystretch}{1.05} 
\begin{tabular}{|c|r|c|r|c|}
\hline
      &\multicolumn{2}{|c}{E835-I}&\multicolumn{2}{|c|}{E835-II}\\ 
\cline{2-5}
 $E_{CM}$ range &  $L_{int}$ & \etacgam 
       & $L_{int}$  & \etacgam \\
   (MeV)  &(pb$^{-1}$)& cand. 
         &(pb$^{-1}$)& cand.
 \\
\hline
 $3300-3400$  &  -    & - &   8.51 & 2  \\
 $3400-3440$  &  -    & - &  21.83 & 3  \\
 $3440-3500$  &  0.50 &   &   2.51 &    \\
\hline
\multicolumn{5}{|c|}{$\chi_{c1}$, $\chi_{c.o.g}$ region and $\chi_{c2}$}\\ 
\hline 
 $3500-3520$     &  5.32 &   &   5.98 &   \\
$3520.0-3525.7$ & 11.24 &   &  13.40 & 1 \\
 $3525.7-3526.2$ & 17.61 & 7 &  11.56 & 6 \\
 $3526.2-3526.7$ &  5.23 & 1 &  17.53 &   \\
 $3526.7-3540.0$ &  8.12 &   &   8.22 &   \\
 $3540-3560$     & 11.10 & 1 &   0.90 &   \\
\hline
 $3560-3675$  & 33.99 & 2 &   0.78 &   \\  
 $3680-3700$  &  9.00 &   &  12.58 &   \\
 $3700-3850$  &  0.98 &   &   6.32 &   \\ 
 $3850-4400$  &  1.63 &   &   2.10 &   \\
\hline
\end{tabular}
\caption{$\eta_c\gamma$ candidates
in each energy interval and corresponding integrated luminosities
{\it ~(Preliminary)}}
\label{data}
\end{center}
\end{table}

The final sample has 23 \etacgam\  candidates; 
15 of them are at  $\chi_{c.o.g}$ (Fig.~\ref{lumplot}). 
Candidates in each energy interval and the corresponding 
integrated luninosities are listed in Table~\ref{data}. 
The observed cross section for \ppbar\ra\etacgam\ is plotted in Fig.~\ref{cross}.
The background is large near 
$E_{CM}=3400\,$MeV but decreases rapidly with energy. 
We performed several checks, in particular: we have analyzed the two data sets
separately and we find compatible excess at the same mass;
If we impose in the fit 
$M_{\eta_c}=2850; 3150\,$MeV, the cross section agrees very well
with smooth background; For data outside the $\chi_{c.o.g.}$
we rescale all energies by 3526.2/$E_{CM}$ and verify that
events would not form an $\eta_c$.

We estimate the significance of the  
excess in the \etacgam\ channel with several methods: 

1. Binomial significance with {\it a priori} hypothesis:
We calculate the cumulative probability for detecting at least $N_{s}$
candidates in an {\it a priori} signal bin, having observed  
$N_{tot}$ candidates in the $E_{CM}$ range that
extends from the $\chi_{c1}$ to $\chi_{c2}$, under the
hyothesis of constant cross section.
As signal bin we take 3525.6$<E_{CM}<$3526.4~MeV where E760 
observed an excess of \psipiz.

2. Binomial significance with {\it a~posteriori} hypothesis
with correction for multiple hypotheses: 
We observe the excess of events in a 0.5 MeV bin between 
$3525.7<E_{CM}<3526.2\,$MeV or a 1.0~MeV bin between 
$3525.7<E_{CM}<3526.7\,$MeV. Since this bin is chosen
{\it a~posteriori}, the significance is estimated from the cumulative binomial 
probability, calculated as above, multiplied by a conservative factor 10
(or 5) for the number of 
possible signal bin choices.
\begin{table*}
\renewcommand{\arraystretch}{1.05} 
\begin{tabular}{|c|c|c|c|c|c|c|c|c|}
\hline
$\Gamma_R$ & $M_R$ & $ \Gamma_{in} B_{out}$ & $\sigma_0$ & 
$b$ & $-\log{\cal L}$ & $\cal {P}$ & $\cal {P}$ & $\cal {P}$\\
(MeV)&  (MeV) & (eV$\times10^3$) & (fb) &(fb/MeV) & & (nominal) & (sim) & (sim)* \\
\hline
  0.5(fixed)  & 3525.83$^{+0.15}_{-0.16}$ & 4.6$^{+ 1.7}_{- 1.5}$ &  
77$^{+30}_{-24}$ & -0.36 & 34.19 & 0.30$\cdot10^{- 3}$ & 
0.98$\cdot10^{- 3}$ & 0.86$\cdot10^{- 3}$ \\
  1.0(fixed)  & 3525.81$^{+0.22}_{-0.23}$ & 5.5$^{+ 2.3}_{- 1.9}$ &  
81$^{+32}_{-25}$ & -0.38 & 35.66 &  1.5$\cdot10^{- 3}$ &
3.16$\cdot10^{- 3}$ & 2.76$\cdot10^{- 3}$\\
\hline
\multicolumn{3}{|l|}{ No resonance:} &156$^{+35}_{-30}$ & -0.74 & 40.72 \\
\cline{1-6}
\end{tabular}
\caption{Fits to the \etacgam\, cross section in the
energy range 3300$<E_{CM} <$4400~MeV. 
$\cal{P}$(nominal) is calculated from
$\chi^2\approx-2\Delta ln \cal L$.  
$\cal{P}$(sim) and $\cal{P}$(sim)* are the probabilities that
$\Delta ln \cal L$  or both $\Delta ln \cal L$ and $\Gamma_{{\bar p} p} B_{\eta_c\gamma}$ 
exceed the experimental values on simulated experiments.{\it ~(Preliminary)}}  
\label{etacgam_fits}
\end{table*}

3. Poisson significance: 
From a linear fit to the background cross section $\sigma_b$ over the full 
energy range $3300<E_{CM}<4400\,$MeV, we estimate 
$\sigma_b(3526.2~MeV)=0.079\pm 0.016\,$pb.
We then estimate the significance from the probability that the expected background
fluctuates to $\geq 13$ events in  $3525.7<E_{CM}<3.526.2\,$MeV, or
$\geq 14$ events in  $3525.7<E_{CM}<3.526.7\,$MeV, multiplied
by 10 (or 5).

4. Significance from likelihood ratio:  
We simulate the outcome of
50,000 experiments under the hypothesis of a
linear background, whose parameters are Gaussian distributed
with mean and variance taken from the  ``no resonance'' fit to the data 
in Tab.~\ref{etacgam_fits}.
For each experiment we perform  
maximum likelihood fits to the null hypothesis $(H0)$ (no resonance) and 
the alternate hypothesis ($H1$) that includes a resonance as described below. 
We then estimate the significance
from the probability that a likelihood ratio 
at least as large as that observed on data can arise by chance.

The most conservative estimate of the significance is obtained
by method 4 which gives $\cal P$ between 1 and
3$\times 10^{-3}$ depending upon the assumed resonance width.
Other methods yield $8\cdot10^{-5}<{\cal P}<3\cdot10^{-3}$.
In the absence of a narrow peaking background, this is strong evidence for a 
resonance near 3526~MeV.

We perform a Poisson maximum
likelihood fit to the measured cross section between
3300 and 4400~MeV as the sum of 
a linearly varying 
background cross section 
($\sigma_{b}(E)=\sigma_0+b(E({\rm
MeV})-3526.2)$)
and a Breit Wigner
convolved with a Gaussian describing the beam energy distribution.

The parameters determined by the fit are 
$\sigma_0$, $b$, $M_R$, and $\Gamma_{in}{\cal B}_{out}=
\Gamma(h_c\to p\bar p){\cal B}(h_c\to \eta_c\gamma)
{\cal B}(\eta_c\to\gamma\gamma)$.
Data are insufficient to fit for $\Gamma_R$ and we perform fits
for fixed values of $\Gamma_R$ between 0.5 and 1~MeV.
The results are given in Table \ref{etacgam_fits} for two 
values of $\Gamma_R$. The background parameters are
relatively independent of $\Gamma_R$, and $\Gamma_{in}{\cal B}_{out}$
only changes by $\approx 20\%$ as $\Gamma_R$ is icreased from 0.5 to
1.0 MeV. The maximum of the likelihood seem to favour smaller $\Gamma_R$.

We also perform a fit including in the model the  
$\propto \sin^2\theta^*_3$ distribution of the E1 radiative transition.
The background angular distribution, measured on a sample enriched with feed-down events 
at the $\chi_{c.o.g.}$ region,
is compatible with isotropy in $-0.4 < \cos \theta^*_3 < 0.7$.
The fitted parameters do not significantly change but $ln \cal L$ increases 
by 2.68
suggesting that the expected angular distribution is a better
hypothesis than isotropic decay.

Dividing $\Gamma_{in}{\cal B}_{out}$ by the value of 
${\cal B}(\eta_c\to\gamma\gamma)=(4.3\pm1.5)10^{-4}\,$\cite{pdg}, we derive  
$10.6\pm3.7\pm3.4(br)<\Gamma_{\overline{p}p}{\cal B}_{\eta_c\gamma}<
12.8\pm4.8\pm4.5(br)\,$eV,  
consistent with ${\cal B}(p\bar p)\approx 1-3\times 10^{-5}$
in the expected range~\cite{kuang} for  
$\Gamma(h_c\to\eta_c\gamma)\approx 600\,$keV.

\subsection{Search for $\bar p p\to h_c\to J/\psi\pi^0\to e^+e^-\gamma\gamma$}
The trigger for $e^+e^-\,X$ final states requires at least two charged tracks 
from the interaction point associated to a signal in the threshold {\v C}erenkov counter
and at least 2 energy deposits in the 
CCAL with invariant mass $\geq 2.2\,$GeV.

\begin{figure}[htb] \begin{center}
\includegraphics[scale=0.42,bb=33 250 550 660]{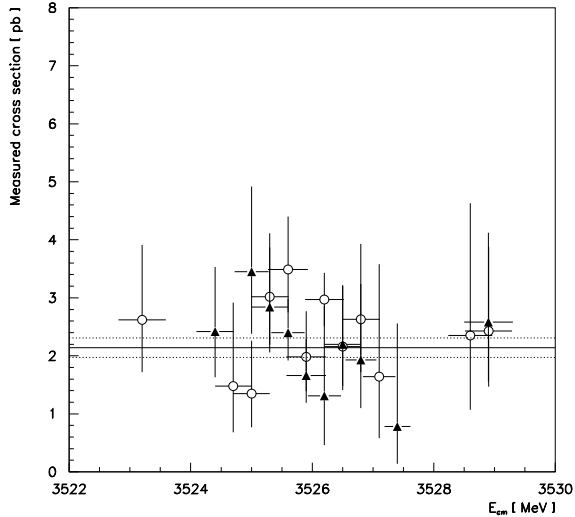}
\end{center}
\caption{Cross section for \ppbar\ra\psipiz\ra$e^+e^-\gamma\gamma$.
E835-I data are open circles and  E835-II data closed
circles {\it ~(Preliminary)}}
\label{cross_psipiz}
\end{figure}

Events must have two electrons
identified by a likelihood ratio test ($EW_1EW_2>1.5\,$\cite{bigpaper}) 
based on $dE/dX$ in the
scintillators, number of ph.e. in {\v C}erenkov and shower 
lateral shape in the CCAL.
We require that the \ep\ invariant mass is $>2800\,$MeV
and limit the $e^+(e^-)$ acceptance to $15^\circ<\theta<60^\circ$.
Acceptance for photons is $11^\circ<\theta<70^\circ$.
We allow additional on-time CCAL
clusters only if compatible with photons radiated by the
$e^\pm$ ($E<100$ MeV and $\theta_{\gamma e} <
10^\circ$). 
Finally the $\chi^2$ probability for the 6C fit to \psipiz\ must exceed 0.01.

We exclude the $\chi_{c1}$ and $\chi_{c2}$ data, since radiative
decays to $J/\psi\gamma$ constitute a background to \psipiz, and
consider only data for 3.52 GeV $<\sqrt s < 3.54$ GeV.
The observed cross section for \ppbar\ra\psipiz\ is plotted in Fig.~\ref{cross_psipiz}. 
Data are compatible with a flat cross section between 3.52 and 3.54 GeV.
Within our acceptance (smaller than that of E760) we see no excess 
of events that would correspond to a 
narrow ($< 1\,$MeV) resonance. 

\section{CONCLUSIONS}
We have measured the cross section for ${\bar p} p \rightarrow \eta_c
\gamma \rightarrow 3\gamma$ near the center of
gravity of the charmonium $^3P_J$ states and observe a narrow structure
($\Gamma \leq 1$ MeV) centered at $3528.8 \pm 0.2\pm0.2\,$MeV. The
statistical significance is ${\cal P}\sim 0.001$. The value
of $\Gamma_{{\bar p} p} B_{\eta_c\gamma}\sim 12.0\pm
4.5\pm4.3\,$eV is compatible with the expected ${\cal B}$(\ppbar) and 
E1 radiative width of the $h_c$. The signal is seen with comparable yield and at the
same $M_R$ in both runs. Our former experiment
E760 had no sensitivity to this channel: with a factor 5 less 
luminosity in this region, it
would have observed between 2 and 3 events.
This observation is evidence for the $h_c$ at its expected location near the
$\chi_{c.o.g}$.
We have measured the cross section for ${\bar p} p \rightarrow J/\psi
\pi^0$ and observe no significant excess in the $\chi_{cJ}$
center-of-gravity region.


\begin{thebibliography}{99}
\bibitem{explan} Event yield at resonance peak for 
reaction~(\ref{psipio})
is estimated from the E760 result,  while
for reaction~(\ref{tregam}) it is estimated assuming 
$\Gamma(h_c\to\gamma\eta_c)\approx 500\,$keV,
${\cal B}(h_c\to p\bar p)\approx 10^{-5}\,$\cite{kuang} and the known
${\cal B}(\eta_c\to\gamma\gamma)\,$\cite{pdg}.
\end{thebibliography}
\end{document}